\documentstyle[twocolumn,prb,aps]{revtex}
\input{epsf.tex}

\begin{document}
\def \beq{\begin{equation}}
\def \eeq{\end{equation}}
\def \beqarr{\begin{eqnarray}}
\def \eeqarr{\end{eqnarray}}

\twocolumn[\hsize\textwidth\columnwidth\hsize\csname @twocolumnfalse\endcsname

\draft
\title{
Wigner Crystals in the lowest Landau level at low filling factors
}

\author{Kun Yang$^1$, F. D. M. Haldane$^2$, and E. H. Rezayi$^3$}
\address{
$^1$National High Magnetic Field Laboratory and Department of Physics,
Florida State University, Tallahassee, Florida 32306
}

\address{
$^2$Department of Physics, Princeton University, Princeton, New Jersey 08544}

\address{
$^3$Department of Physics, California State University, Los Angeles,
California 90032}

\date{May 8, 2001}
 
\maketitle
\begin{abstract}
We report on results of finite-size numerical studies of partially
filled lowest Landau level, at low electron filling factors. We find convincing
evidence suggesting that electrons form Wigner Crystals at sufficiently low
filling factors, and the critical filling factor is $\nu_c\approx 1/7$. At
$\nu= 1/7$ we find the system undergoes a phase transition from the Wigner 
Crystal to the incompressible Laughlin state when the short-range part of the
Coulomb interaction is modified slightly. This transition is either continuous
or very weakly first order.
 
\end{abstract}

]

Interacting electrons at sufficiently low 
densities and temperatures were shown by Wigner\cite{wigner} to form a crystalline state.
This ground state results from the dominance of the potential over the kinetic energy  which 
occurs in the dilute limit.
A Wigner Crystal (WC) state has been observed in
a low-density {\em two-dimensional} (2D) electron gas system trapped on the 
surface of a helium liquid.\cite{helium} While unlikely to be observed at metallic 
densities in 3D this crystalline phase is in fact expected to occur 
in the 2D electron gas confined in semiconductor heterostructures or
quantum wells, and subject to a strong perpendicular magnetic field (such that all
electrons are in the lowest Landau level (LLL)). Under these conditions the 
kinetic energy is completely quenched making it an ideal environment in which to
observe Wigner crystallization;
the WC state is expected to have the lowest Coulomb interaction energy when
the Landau level filling factor ($\nu$) is sufficiently low. Indeed, substantial
experimental evidence, mostly from transport measurements, points to a 
pinned WC state in the insulating phases of low $\nu$.\cite{review} However the
situation is not yet definitive as  direct experimental probe of 
the lattice structure of the WC is still lacking. In addition,  the ubiquitous 
disorder in these systems could lead to other insulating phases\cite{review} such as the Hall insulator, or
just complicate the detection of the crystal itself. On the theoretical side, early
work\cite{anderson} suggested
that a 2D electron gas is always unstable against formation of a WC when 
subject to a strong magnetic field, at zero temperature. However,
the discovery of fractional quantum Hall
effect (FQHE) has led to the realization\cite{laughlin} 
that  electrons  can also form 
incompressible liquid states, especially at the primary sequence 
$\nu=1/m$, where $m=3, 5, \cdots$.
By comparing the energy of the Laughlin\cite{laughlin} state with that of a 
{\em correlated} WC state (whose energy is considerably lower than that of a 
simple Hartree-Fock estimate\cite{yoshioka,maki}), 
Lam and Girvin\cite{lam} concluded that the critical
filling factor $\nu_c$ below which the WC forms is slightly above $1/7$.
This is in good agreement with transport experiments on clean samples.
A more recent study\cite{jain} 
using a formalism based on composite fermion\cite{jain0} (CF) wavefunctions
calculated the collective excitation energy variationally and 
found that the ground state become unstable 
against proliferation of magneto-rotons (which points to formation of WC)
for $\nu\le 1/9$, although such an instability was not found at $\nu=1/7$.
A similar trend was observed earlier in the single mode approximation\cite{GMP} 
where the magneto-roton gap while not zero was found to be diminished 
considerably at 1/7 and further at 1/9. This was interpreted as a precursor to magneto-roton 
mode softening and the transition to the WC phase. 

In this paper we present finite size exact diagonalization studies of systems
with low $\nu$ which directly reveal the formation of the WC for the first time in the 
LLL. 
We study systems with torus geometry. Our previous work on 
charge density wave (CDW) ground states in high Landau levels\cite{rezayi,haldane}
has demonstrated that the torus geometry is advantageous as compared to, say, 
the spherical geometry\cite{haldanesph}, in studies of states
with broken translational symmetry. The reasons are: (i) on a torus one can 
adjust the geometry to better accommodate a lattice in a finite-size system.
(ii) The presence of translational symmetry allows one to define a many-body 
momentum\cite{duncan} that can be used to label the eigenstates. By inspecting
the momenta of the low-lying 
states, as well as the wavevector dependence of the density-density correlation
functions, one can easily detect whether the translational symmetry 
is broken in the system, and determine lattice structure  
of the broken symmetry (crystal) state. 

We find strong evidence that WC forms at filling factors $\nu\le 1/7$, and that
the critical filling factor $\nu_c$ is very close to $1/7$. Furthermore,
we have studied the competition between the WC state and the Laughlin state
at $\nu=1/7$ by varying the geometries of our finite-size systems, as well as
the short-range part of the Coulomb interaction. We find that the Laughlin 
state is always {\em continuously} deformed into one of the low-lying states 
that correspond to the WC phase, when the geometry or the interaction is varied.
This suggests that the quantum phase transition between the incompressible 
FQHE and compressible WC phases is either continuous or very weakly first-order.
The methods used here are similar to our previous studies and provide 
a reliable way of detecting broken symmetry in finite systems.   We
first investigate dependence  of the spectrum on PBC geometry.

\begin{figure}
\vspace{2truecm}
\centerline{
\epsfxsize=8cm
\epsfbox{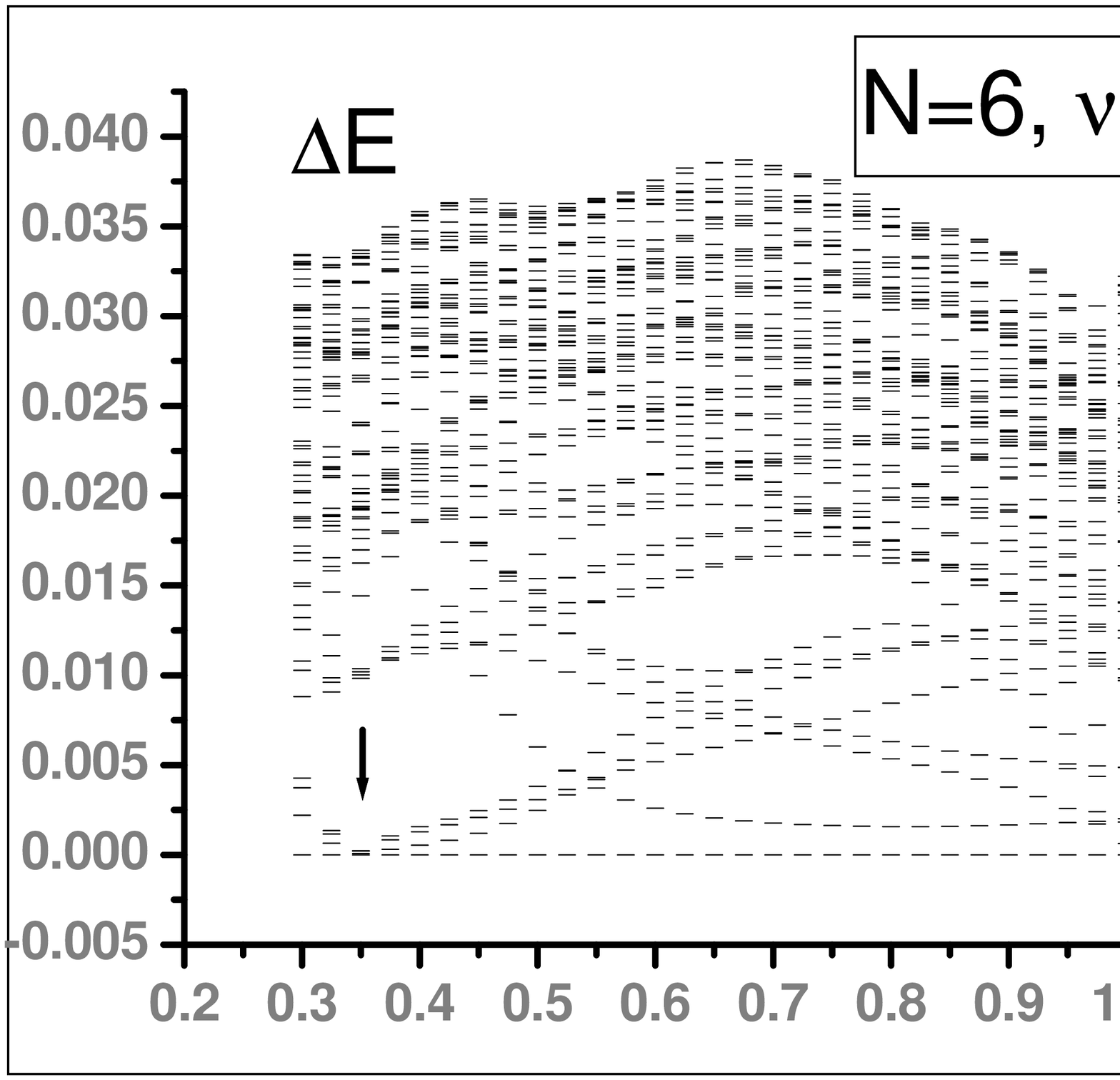}
}
\centerline{
\epsfxsize=8cm
\epsfbox{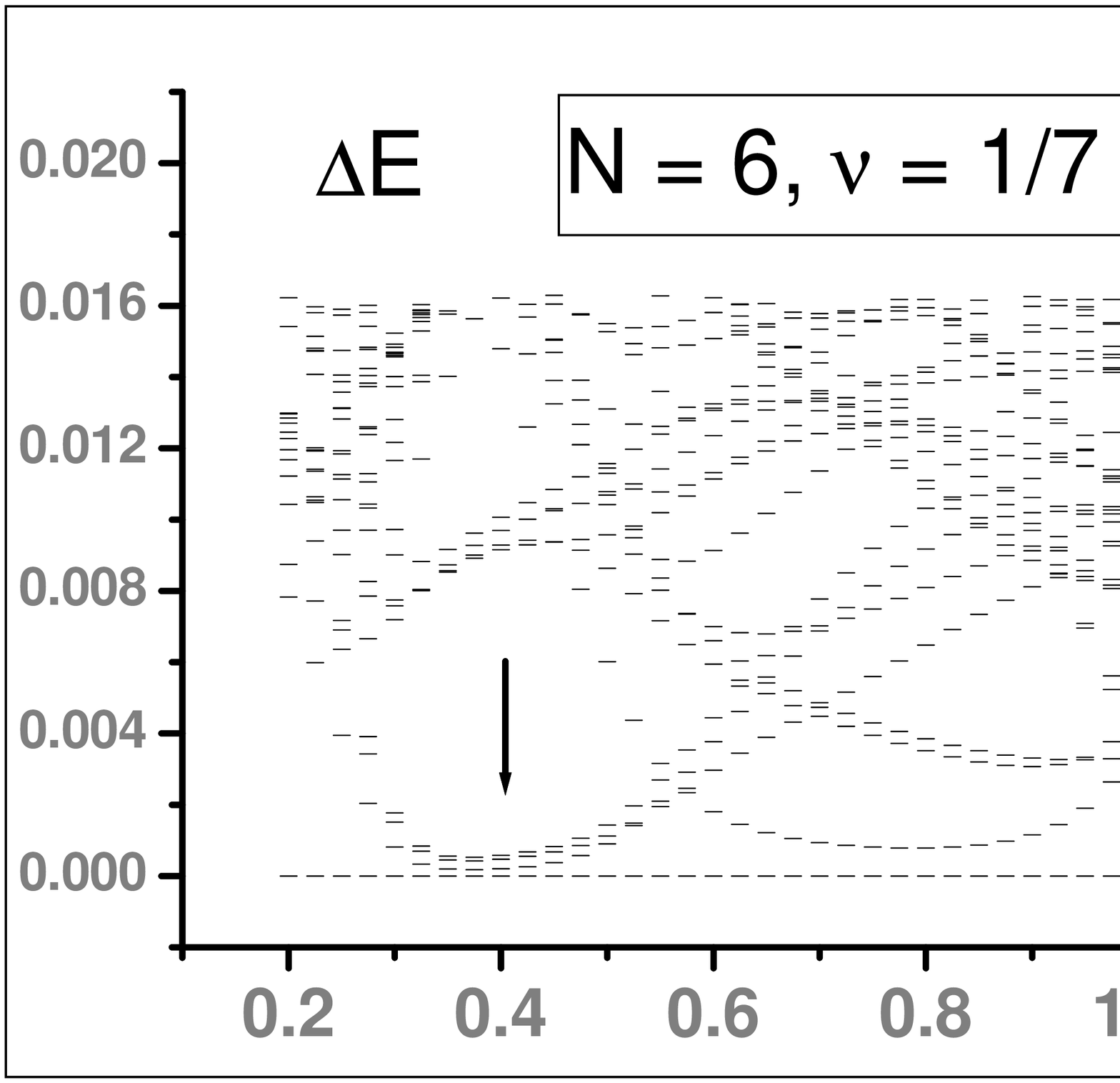}
}
\centerline{
\epsfxsize=8cm
\epsfbox{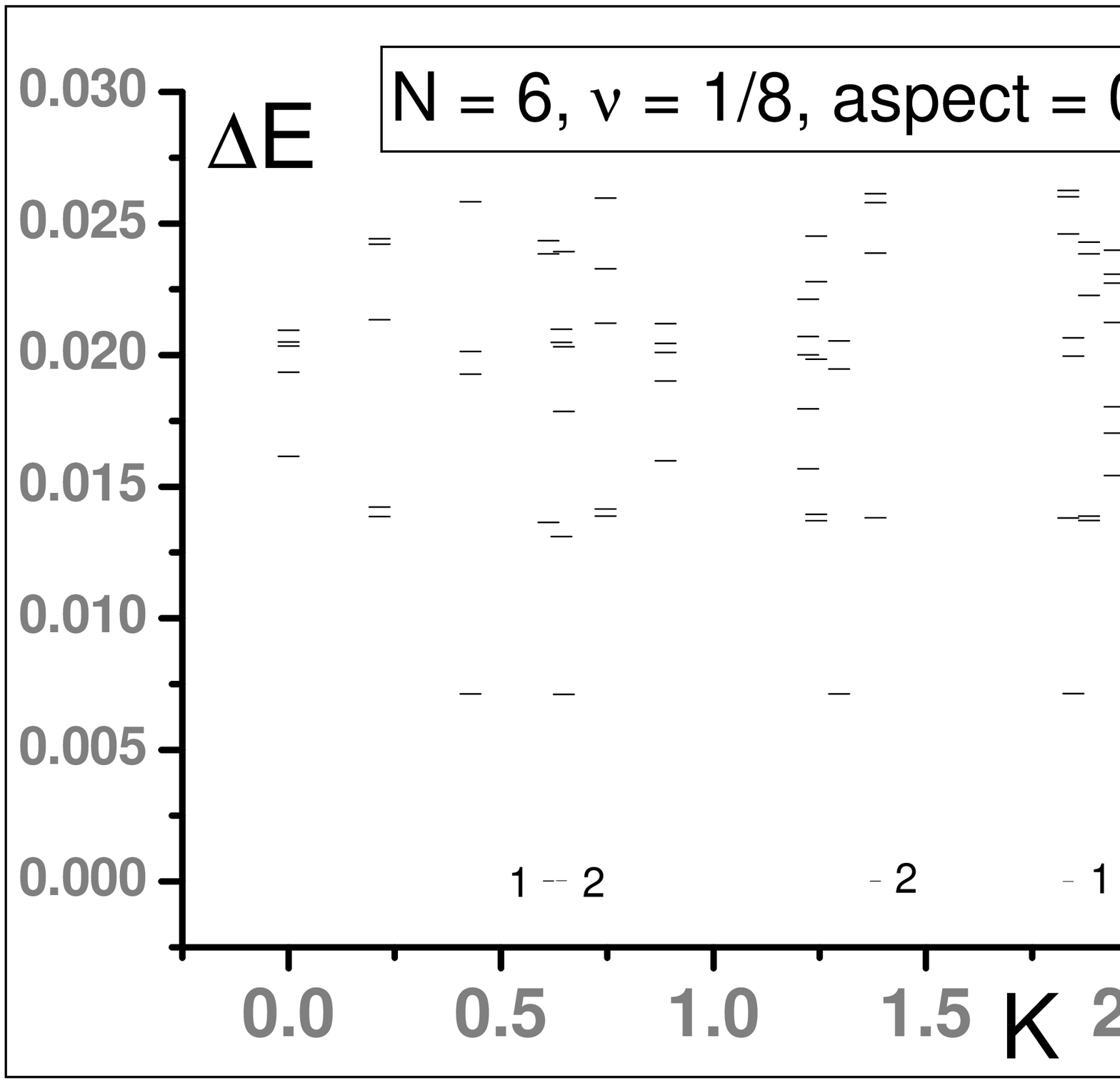}
}
\vspace{-1.5truecm}
\caption{
The spectra of rectangular-shaped finite-size systems with $N=6$ electrons, at
filling factors 1/6, 1/7, and 1/8. The energy levels are plotted as functions of
the aspect ratio for $\nu=1/6$ and 1/7, and plotted as a function of the 
magnitude of momentum ${\bf K}$ for $\nu=1/8$, at aspect ratio 0.35. The 
degeneracy of each low-lying level is indicated by the integer beside it for
$\nu=1/8$. 
}
\label{fig1}
\end{figure}

Fig. 1 shows the energy spectra of systems with $N=6$ electrons, at
filling factors $\nu=1/6, 1/7$, and $1/8$, with rectangular geometry, for
pure Coulomb interaction.
For $\nu=1/6$ and $1/7$ we plot 
energy levels for a series of aspect ratio $a_0$, while for $\nu=1/8$ we
plot the energy level versus the magnitude of many-body momentum ${\bf K}$,
for $a_0=0.35$. One can see that for the $\nu=1/8$ case, as well as
$\nu=1/7$ for a sizable range of $a_0$: $0.3 < a_0 < 0.5$, there are $N_D=6$ nearly
degenerate low-lying states (including the ground state) that form what we
call the ground state manifold.
The ${\bf K}$'s of the states in this manifold form a 2D array, as we show
in Fig. 2 for $\nu=1/7$ at $a_0=3/8$, indicating
that the translational symmetry is broken in both directions, and the system
has 2D crystalline order. These ${\bf K}$'s  
determine the
lattice structure
of the crystal. The number of unit cells included in the finite size
system can be
determined easily\cite{haldane}:
$N_c=\overline{N}^2/N_D$, where
$\overline{N}$ is the highest common divisor of $N$ and the number of flux
quanta $N_\phi$. Here we have $\overline{N}=6$ in both cases, thus
$N_c=6=N$, indicating there is one electron per
unit cell. This is precisely 
what one expects for a Wigner Crystal (WC). This suggests that electrons form
a WC at these filling factors, at least when a WC can be easily accommodated
by the geometry of the finite size system. 
\begin{figure*}[h]
\vspace{2truecm}
\centerline{
\epsfxsize=9cm
\epsfbox{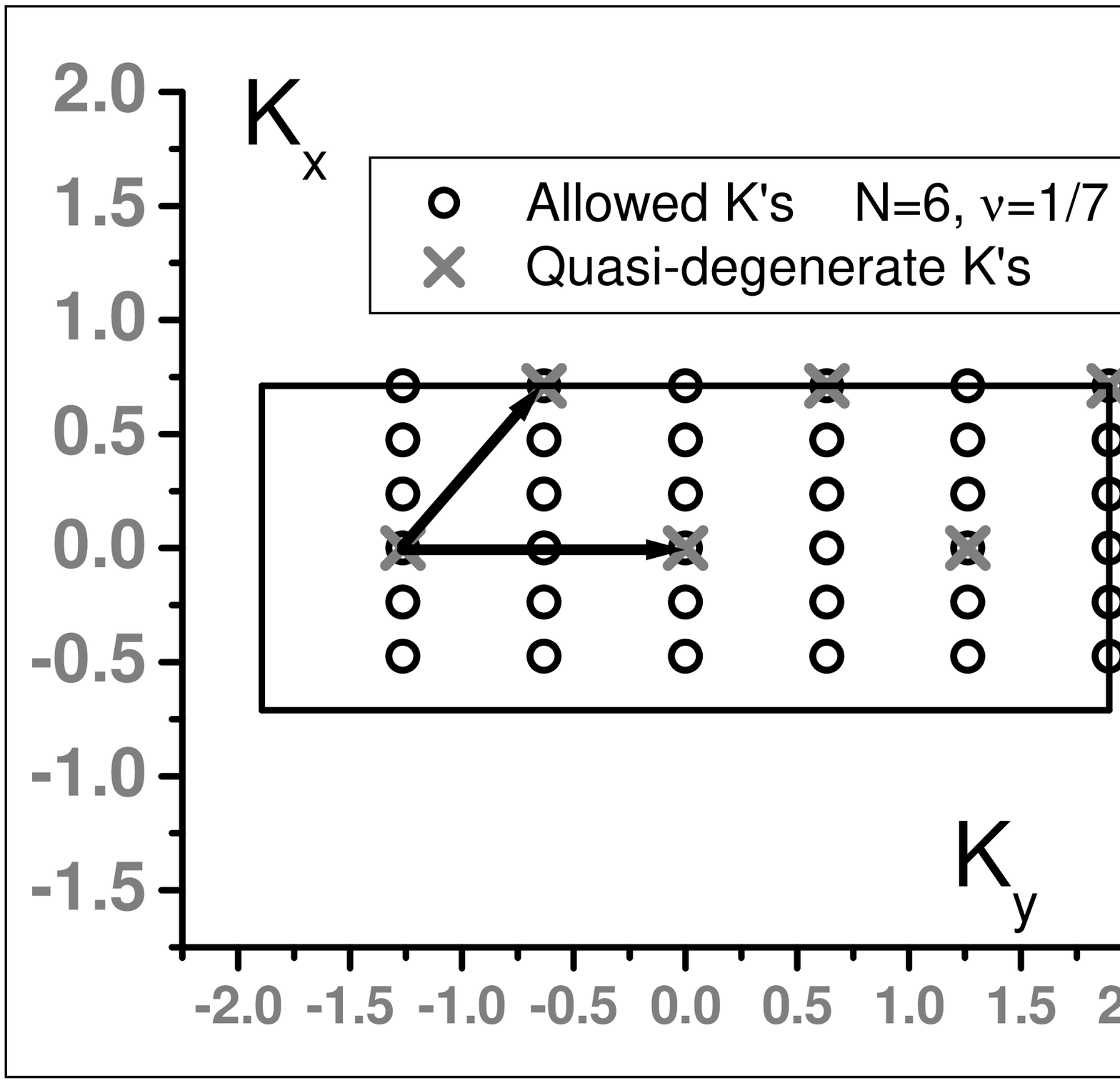}
}
\vspace{-1.5truecm}
\caption{
The allowed momenta ${\bf K}$ for the $\nu=1/7$ system with $N=6$ electrons
with aspect ration $a_0=3/8$, and the ${\bf K}$'s of the ground state manifold.
These low-lying ${\bf K}$'s form a nearly triangular array, which
represent the reciprocal lattice of the Wigner Crystal that the electrons form.
}
\label{fig2}
\end{figure*}
\noindent
For $\nu=1/6$, on the other hand,
such a (near) degeneracy is absent for almost all geometries, except for 
perhaps at $a_0\approx 0.35$. This suggests that the translational symmetry is 
not broken at $\nu=1/6$, although the tendency toward WC formation is present,
especially when the geometry of the finite-size system is favorable. We thus
conclude that the critical filling factor $\nu_c$ below which the Wigner 
Crystal forms is close to and probably slightly above $1/7$, which is in 
excellent agreement with the estimate of Lam and Girvin.\cite{lam}
>From the superlattice primitive basis vectors in the reciprocal space\cite{haldane} 
we obtain the 
parameters of the WC unit cell $a_1\approx 4.421$, $a_2\approx 6.655$, and 
$\theta\approx 131.6^\circ$.  While
this is not exactly a triangular lattice it is close to it.  We believe the discrepancy
is caused 
by the finite size of the system, and in the thermodynamic limit we expect 
to recover
a triangular lattice. We emphasize the fact that the WC state is stabilized 
near $a_0=0.4$ is {\em not} an artifact of the geometry of the finite size
system; as we will see below at filling factors $\nu=1/3$ and 1/5 where the 
system is an incompressible fluid and well described by the Laughlin state, no 
crystalline order is developed and the ground state has high overlap with the
Laughlin state for all reasonable geometries, including the ones that are
favorable for WC formation.

\begin{figure}
\vspace{2truecm}
\centerline{
\epsfxsize=9cm
\epsfbox{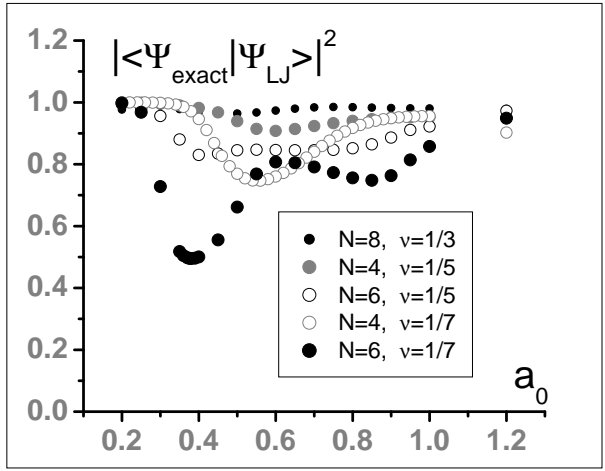}
}
\vspace{-1.5truecm}
\caption{
The square of the overlap between the Laughlin state and the exact ground state
of systems with rectangular geometry and a variety of filling factors and 
number of electrons, as a function of aspect ratio. The points at 1.2 are for
hexagonal unit cells.}
\label{fig3}
\end{figure}

At $\nu=1/m$ with $m$ being an odd integer, it is known\cite{book} that for certain
short-range repulsive interactions, the 
Laughlin state is the exact ground state that describes an incompressible 
fractional quantum Hall liquid. In Fig. 3 we show the square of the overlap 
between the Laughlin state with the exact ground state of the system for 
$m=3, 5$ and 7, at different aspect ratios and system sizes. It is seen that
for $\nu=1/3$ and $1/5$, the overlap is rather insensitive to the geometry of the unit cell and
is close to 1, 
while for $\nu=1/7$ the square of the 
overlap varies appreciably with the geometry, 
and dips below 0.75 and 0.5 in certain
range of $a_0$ for $N=4$ and $N=6$ respectively. The  sensitivity of the 
energy spectrum as well as the overlaps on the system geometry 
is a strong
indication that the ground state is {\em compressible}. It reflects the fact that the
system wants to form a WC, but is frustrated when the aspect ratio $a_0$ is far
from the optimal one for forming a crystal, 
$a_0\approx 0.4$
for $N=6$. It should be noted that the
Laughlin state has ${\bf K}=0$, which is one of the ${\bf K}$'s in the
ground state manifold, and in the case of a pure Coulomb interaction the ground 
state turn out to be in the ${\bf K}=0$ sector for all the aspect ratios and sizes that we
explored; thus the square of the
overlap evolves continuously from being close to 1 near $a_0=1$, where the WC 
state is frustrated,
to be less than 0.5 near $a_0=0.4$ where there is little geometric frustration 
for WC formation. Physically the reason that the overlap is large when the
WC state is frustrated is that the Laughlin state at $\nu=1/7$, 
while describing an
incompressible liquid state, already has substantial short-range crystal order
built into it;
it is thus energetically still competitive.

\begin{figure}
\vspace{2truecm}
\centerline{
\epsfxsize=9cm
\epsfbox{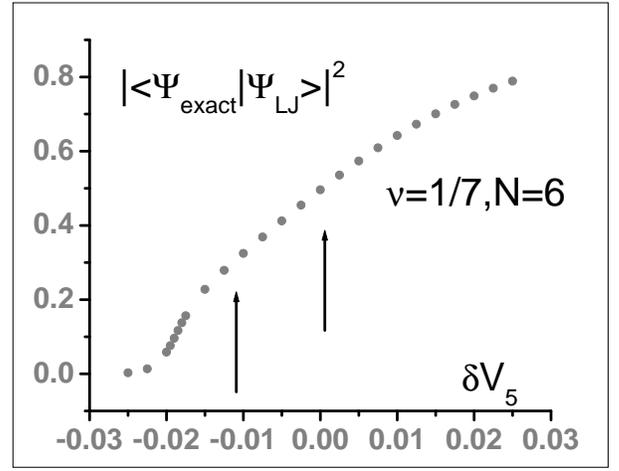}
}
\vspace{-1.5truecm}
\caption{
The square of the overlap between the Laughlin state and the exact ground state,
in a system with 
aspect ratio $a_0=3/8$. It is plotted as a function of $\delta V_5$ 
(see text). The arrow on the right hand side indicates the position
at which the near-degeneracy of the low-lying states starts to develop.
The arrow on the left hand side indicates the position of $\delta V_5$ 
below which the absolute ground state is not in the $K=$ sector. 
The overlap shown 
is between the Laughlin state and the lowest energy $K=0$ state.
}
\label{fig4}
\end{figure}

At $\nu=1/7$, the Laughlin state can be stabilized by slightly modifying the
short range part of the Coulomb potential. 
In Fig. 4 we plot the square of the overlap between the Laughlin  
state and the exact ground state of the system at $a_0=3/8$, as a function of 
$\delta V_5$, the change of the $m=5$ pseudo-potential\cite{book} from its Coulomb value.
It is seen that the 
overlap is substantially improved to be above 0.8 by adding a small 
pseudo-potential $\delta V_5=0.025e^2/\epsilon\ell$ (which represents less than  $10\%$
increase of the $V_5$ pseudo-potential), even though the system geometry is most
favorable for WC formation. This is further indication that $\nu=1/7$ is very 
close to the phase boundary separating the crystal and liquid phases. The 
smooth change of the overlap as a function of $\delta V_5$ also suggests the 
transition is either continuous or very weakly first-order in the thermodynamic
limit. Adding a negative $\delta V_5$, on the other hand, 
further suppresses the overlap; in particular, at $\delta V_5\approx -0.011
e^2/\epsilon\ell$ there is a level crossing for the ground state, and 
for $\delta V_5 < -0.011e^2/\epsilon\ell$ the {\em absolute} ground state has a 
 {\em different} wavevector from the Laughlin state; thus the
overlap between the Laughlin state and the absolute ground state is strictly 
speaking zero
here.  However, the quasi-degeneraces are well developed 
and the ground state is not unique.  We have tracked the overlap with the $K=0$ ``ground
state''. In real systems, the short-range part of the Coulomb interaction is
{\em softened} due to the finite extent of the electron wave function along the
$\hat{z}$ direction (or finite layer thickness), 
thus the WC state is expected to be further stabilized; we have studied
potentials of typical layer thickness and
find it is indeed the case at $\nu=1/7$.

\begin{figure}
\vspace{2truecm}
\voffset 1.8in
\centerline{
\epsfxsize=9cm
\epsfbox{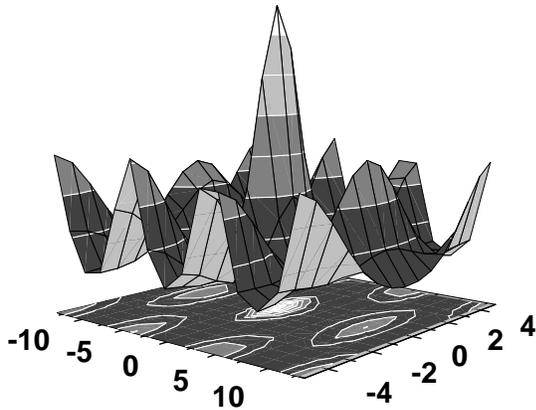}
}
\vspace{-1.5truecm}
\caption{
The real-space lowest Landau level projected 
(or guiding center) density-density 
correlation function, for $N=6$ electrons at filling factor 1/7, in 
a rectangular system with aspect ratio 3/8. 
A 2D contour plot is also shown.
}
\label{fig5}
\end{figure}

As in our previous studies we now turn to the density response functions.
In Fig. 5 we show the LLL projected (or guiding center) density-density
correlation function, in real space. This is the Fourier transform of the
guiding center static structure factor\cite{haldane}. The central peak represents the
usual $\delta({\bf r})$ singularity projected to the LLL, which is at the position
of the electron where the correlation is
measured from. The crystal structure can be clearly seen; the number of
peaks (including the central one) equals the number of electrons $N=6$,
indicating each unit cell includes one electron, which is what one expects
for a Wigner Crystal.

The tendency toward WC formation, as well as the structure of the resultant
lattice, can also be seen in the density response function in momentum space,
$\chi({\bf q})$. Fig. 6 is a plot of $\chi({\bf q})$ for the same system as
in Fig. 5. Here we see strong response at an array of momenta, which form a
lattice that is close to being triangular. This is essentially the
reciprocal lattice of the crystal that forms in real space.
The origin of the large response at these
momenta is the near-degeneracy of the low-lying states that are connected by
these momenta discussed earlier; small energy denominators lead to large
response at these momenta.

\begin{figure}
\vspace{1.7truecm}
\centerline{
\epsfxsize=9cm
\epsfbox{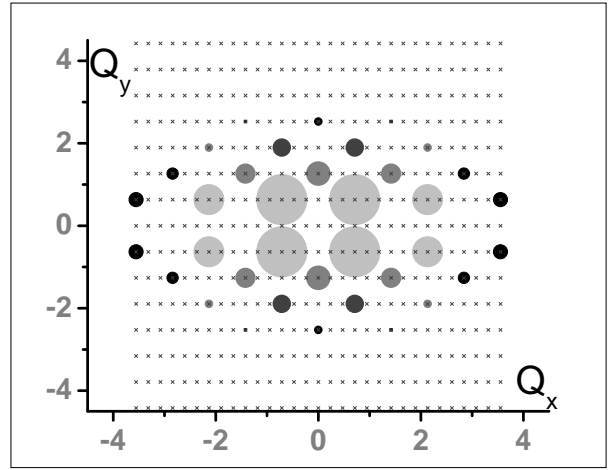}
}
\vspace{-1.5truecm}
\caption{
A 2D plot of the LLL projected density response function $\chi({\bf q})$, with
the size of the dot representing the strength. The system is the same as that
if Fig. 5.  The 4 large dots represent a peak value of about 21000.  The 
nearby smaller dots along the $Q_y$ axis represent a value of 4700 etc.
}
\label{fig6}
\end{figure}

In summary, we have shown that FQHE incompressible fluids become unstable around
$\nu=1/7$ in agreement with previous predictions. We find the Wigner crystal 
becomes the ground state as the system geometry is adjusted to accommodate the
crystal.  It is more difficult to determine the order of the transition by the
small sizes that we have studied.  However, a strong level crossing first order transition 
seems to be ruled out by our study.  We observe only a continuous evolution
of the required quasi-degenerate manifold which always includes the state
at $K=0$.  This state for geometries that frustrate the crystal shows large
overlap with the Laughlin's state. This suggest the transition may be second order
but a weakly first order transition predicted by a mean field approach\cite{GMP} also
remains a possibility.

FDMH and EHR thank the Aspen Center for Physics where this work was started. EHR
is grateful to NHMFL for their hospitality where it was brought to a conclusion.
This work
was supported by NSF grants DMR-9809483 (FDMH), DMR-0086191 (EHR),
DMR-9971541 (KY), and the A. P. Sloan Foundation (KY).

\end{document}